\documentclass{article}

\usepackage{microtype}
\usepackage{graphicx}
\usepackage{subfigure}
\usepackage{booktabs} 

\usepackage{hyperref}

\usepackage[accepted]{icml2024}

\usepackage{amsmath}
\usepackage{amssymb}
\usepackage{mathtools}
\usepackage{amsthm}

\usepackage[capitalize,noabbrev]{cleveref}

\theoremstyle{plain}

\theoremstyle{definition}

\theoremstyle{remark}

\usepackage[textsize=tiny]{todonotes}
\usepackage{amsmath}
\usepackage{graphics}
\usepackage{epsfig}
\usepackage{multirow}
\usepackage{xspace}
\usepackage{subfigure}
\usepackage{fancyhdr}
\usepackage{microtype}
\usepackage{color}
\usepackage{colortbl}
\usepackage{xcolor}
\definecolor{mygray}{gray}{.85}
\definecolor{myyellow}{RGB}{204,102,0}
\definecolor{myred}{RGB}{204,0,102}
\definecolor{mypurple}{RGB}{102,0,204}
\definecolor{maroon}{cmyk}{0,0.87,0.68,0.32}
\definecolor{myblue}{RGB}{227,227,240}
\usepackage{amsmath}
\usepackage{graphics}
\usepackage{epsfig}
\usepackage{multirow}
\usepackage{xspace}
\usepackage{subfigure}
\usepackage{fancyhdr}
\usepackage{amssymb}
\usepackage{amsthm}
\usepackage{mathrsfs}
\usepackage{booktabs}
\usepackage{arydshln}
\usepackage{dashrule}
\usepackage{enumitem}

\usepackage{algorithm}
\usepackage{algorithmic}
\renewcommand{\algorithmicrequire}{\textbf{Input:}} 
\renewcommand{\algorithmicensure}{\textbf{Output:}} 
\newcommand{\model}{SurfPro\xspace}

\icmltitlerunning{\model: Functional Protein Design Based on Continuous Surface}

\begin{document}

\twocolumn[
\icmltitle{\model: Functional Protein Design Based on Continuous Surface}



\icmlsetsymbol{equal}{*}

\begin{icmlauthorlist}
\icmlauthor{Zhenqiao Song}{yyy}
\icmlauthor{Tinglin Huang}{xxx}
\icmlauthor{Lei Li}{yyy}
\icmlauthor{Wengong Jin}{zzz}
\end{icmlauthorlist}

\icmlaffiliation{yyy}{Language Technologies Institute, Carnegie Mellon University, Pittsburgh, the United States.}
\icmlaffiliation{xxx}{Yale University, New Haven, United States.}
\icmlaffiliation{zzz}{Broad Institute of MIT and Harvard, Boston United States}

\icmlcorrespondingauthor{Zhenqiao Song}{zhenqiaosong@cmu.edu}

\icmlkeywords{Machine Learning, ICML}

\vskip 0.3in
]



\printAffiliationsAndNotice{}  

\begin{abstract}
How can we design proteins with desired functions? We are motivated by a chemical intuition that both geometric structure and biochemical properties are critical to a protein's function. In this paper, we propose \model, a new method to generate functional proteins given a desired surface and its associated biochemical properties. \model comprises a hierarchical encoder that progressively models the geometric shape and biochemical features of a protein surface, and an autoregressive decoder to produce an amino acid sequence. We evaluate \model on a standard inverse folding benchmark CATH 4.2 and two functional protein design tasks: protein binder design and enzyme design. Our \model consistently surpasses previous state-of-the-art inverse folding methods, achieving a recovery rate of 57.78\% on CATH 4.2 and higher success rates in terms of protein-protein binding and enzyme-substrate interaction scores.


\end{abstract}

\section{Introduction}
\label{introduction}

\begin{figure}
\begin{minipage}[t]{0.45\linewidth}
\centering
\includegraphics[width=3.7cm]{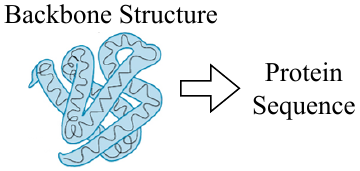}
\centerline{(a) Inverse folding}
\end{minipage}%
\begin{minipage}[t]{0.53\linewidth}
\centering
\includegraphics[width=4.25cm]{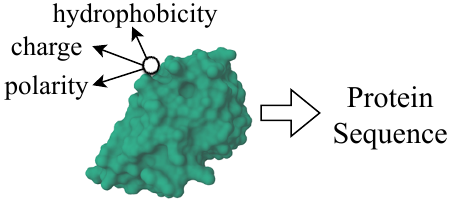}
\centerline{(b) Surface based design}
\end{minipage}
\vspace{-0.6em}
	\caption{ Problem setups of protein design. (a) Inverse folding: protein design conditioned on geometric constraints only. (b) Surface based design: protein design conditioned on both geometric shape and biochemical properties.} 
 \label{Fig: figure_surface_structure}
\end{figure}

Proteins serve diverse functions crucial to cellular processes in our biological system.
In recent years, the remarkable achievements of generative AI have transformed the field of protein design~\cite{huang2016coming,rives2021biological,watson2023novo}. 
One prevalent approach involves first choosing or designing a target backbone structure and then identifies a sequence that folds into this backbone~\cite{dauparas2022robust,anishchenko2021novo,wang2022scaffolding,yeh2023novo}. The first step specifies the geometry of the desired protein (without amino acid types) and the second step (also known as inverse folding, as illustrated in Figure~\ref{Fig: figure_surface_structure} (a)), determines the amino acid composition corresponding to the given shape.

However, the goal of protein design goes beyond predicting a sequence that folds into a target backbone~\cite{defresne2021protein}.
The ultimate goal is to design proteins with desired functions, such as enzymes binding to specific substrates or proteins inhibiting given targets.
The limitation of inverse folding is that it only specifies geometric constraints through the given backbone structure.
To dictate the desired functions, we need to impose not only geometric constraints but also biochemical property constraints.
For example, two proteins with complementary shapes may still not bind due to poorly placed charges, polarity, or hydrophobicity at their binding interface~\cite{gainza2023novo}.

To address this issue, we propose \model, a method to design functional proteins given a biochemical property augmented point cloud~(also called surface based design, Figure~\ref{Fig: figure_surface_structure} (b)).
Each point on the surface is labeled with a three-dimensional~(3D) coordinate and a set of biochemical properties. \model generates an amino acid sequence based on the surface's geometric shape and biochemical properties.
\model comprises a hierarchical encoder and an autoregressive decoder. The encoder progressively models the geometric and biochemical features of the surface through a series of local graph convolutions, followed by global self-attention layers that focus on modeling long-range interactions. The decoder generates a protein sequence based on the learned geometric and biochemical representations of the surface, with the goal that the generated sequence folds into the given surface.

Our contributions are listed as follows:
\begin{itemize}[nosep,leftmargin=2.6em]
\item We propose \model to design functional proteins based on continuous surfaces augmented with biochemical properties.  
\item We evaluate \model on a standard inverse folding benchmark CATH 4.2. \model achieves 57.78\% sequence recovery rate and 3.13 perplexity, significantly outperforming previous inverse folding methods including ProteinMPNN~\cite{dauparas2022robust}, PiFold~\cite{gao2022pifold}, and LM-DESIGN~\cite{zheng2023structure}.
\item We setup a binder design task, and we use AlphaFold2~\cite{jumper2021highly} pAE\_interaction~\cite{bennett2023improving,watson2023novo} to evaluate the binding of designed proteins. \model exhibits superior capability of designing binders with stronger interaction with target proteins than experimentally confirmed positive binders, with an average success rate of 26.22\% across six targets, outperforming the best prior method by 6.9\%.
\item We setup an enzyme design task, and we use ESP score~\cite{kroll2023general,kroll2023turnover} to measure the binding between designed enzymes and their substrates. \model is able 
to design enzymes with higher enzyme-substrate interaction scores than natural enzymes, achieving an average success rate of 43.46\% across five enzyme datasets, outperforming the best prior method by 2.98\%.
\end{itemize}

\section{Related Work}
\label{related_work}
\textbf{Methods for Protein Sequence Design.}
Protein sequence design has been studied with a wide variety of methods. Most studies in this field have adopted one of the three main paradigms: (i) guided by fitness landscape~(function scores), (ii) conditioning on a fixed backbone structure, and (iii) finetuned from a pretrained model on large-scale data.

Protein sequence design guided by fitness landscape includes traditional directed evolution \cite{arnold1998design,dalby2011strategy,packer2015methods,arnold2018directed} and machine learning methods. The mainly used machine learning algorithms include reinforcement learning \cite{angermueller2019model,jain2022biological}, Bayesian optimization \cite{moss2020boss,terayama2021black} and search using deep generative models \cite{brookes2018design,brookes2019conditioning,kumar2020model,ren2022proximal,song2023importance}. Protein design based on a fixed backbone structure is also called inverse folding~\cite{fleishman2011rosettascripts,ingraham2019generative,hsu2022learning,gao2022pifold,zheng2023structure}, which ensures the preservation of a stable structure while also allows for the sampling of diverse sequences.
Due to the available massive sequence data, recent studies have successfully employed machine learning models pre-trained on such data, including ESM~\cite{rives2021biological}, ProtGPT2~\cite{ferruz2022protgpt2} and ProGen~\cite{madani2023large}. This approach has led to notable advancements in addressing various downstream tasks within the field of protein design, such as protein mutation~\cite{meier2021language} and structure-informed protein design~\cite{zheng2023structure}.

\begin{figure*}
\begin{minipage}[t]{0.5\linewidth}
\centering
\includegraphics[width=11.0cm]{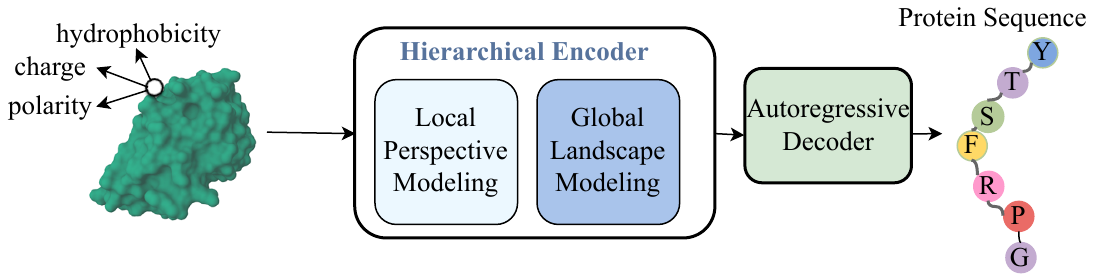}
\centerline{(a) \small{model overview}}
\end{minipage}%
\begin{minipage}[t]{0.65\linewidth}
\centering
\includegraphics[width=5.3cm]{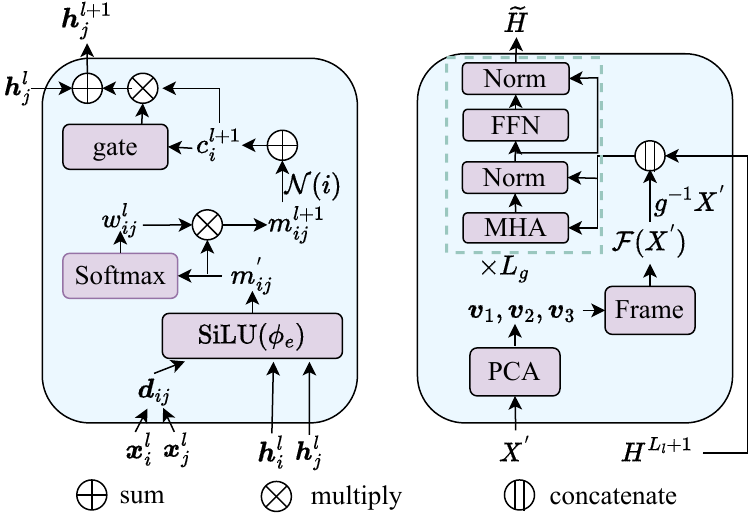}
\centerline{(b) \small{left: local modeling, right: global modeling}}
\end{minipage}
	\caption{(a) The overview of our proposed \model. (b) Left: local perspective modeling, right: global landscape modeling.} 
 \label{Fig: model}
\end{figure*} 

\textbf{Protein Surface Modeling.} Protein design based on its surface is an under-explored area, and most existing studies have not taken into account the significant role of protein molecular surface plays in various biological processes. Traditionally, molecular surfaces are defined using Connelly surfaces~\cite{connolly1983solvent,sanner1996reduced} based on van der Waals radii, often represented as mesh-based structures derived from signed distance functions. Seminal work for modeling protein molecular surfaces is MaSIF~ \cite{gainza2020deciphering}, which fingerprints molecular surfaces expressed as molecular meshes using pre-defined and pre-calculated physical and geometrical features. To remove the high pre-computation costs of featurization, \citet{sverrisson2021fast} propose dMaSIF, showing that modeling molecular surface as a point cloud with atom categories per point is competitive to mesh-based methods. However, both works target at protein understanding tasks, such as protein-protein interactions, instead of protein design.
\citet{gainza2023novo} then expands MaSIF to enable de novo binder design. Initially, they utilize the generated surface fingerprints to predict target binding sites. Following this, they search for binders containing complementary structural motifs, which are subsequently transplanted to protein scaffolds. In our work, we develop a generative model to directly generate functional proteins from their surfaces, eliminating the need for handcrafted feature calculation.


\section{Proposed Method: \model}
\label{methods}
A molecular surface defines the shape of a protein in 3D Euclidean space and the biochemical properties, such as hydrophobicity and charge. The surface shape and the associated biochemical properties co-determine the underlying protein functions. Given a desired surface with geometric and biochemical constraints, how can we generate protein sequences fitting the surface? In this section, we introduce \model, a new functional protein design method based on protein surfaces. Our method works on successive geometric representations of a protein. \model consists of a hierarchical encoder that progressively model the 3D geometric shape and the biochemical features from a local perspective to a global landscape, and an autoregressive decoder that generates a protein sequence based on the geometric and biochemical constraints of the corresponding surface. Figure~\ref{Fig: model} (a) gives an overview of \model.


\subsection{Surface Generation}
\label{sec_surface_generation}
Our method works on successive point clouds of protein surfaces. A high-quality surface should satisfy the following two properties: (1) Smooth: The surface defined by the point cloud should exhibit sufficient smoothness; 
(2) Compact: The point cloud should remove redundant information by down-sampling to improve efficiency.

\textbf{Raw Surface Construction.} We use MSMS~\cite{ewing2010msms} to compute the raw molecular surface of a protein, which is provided as a point cloud with $N$ vertices $\{\boldsymbol{x}_1, \boldsymbol{x}_2, ..., \boldsymbol{x}_N\} \in \mathbb{R}^{N\times 3}$. Suppose the protein is a $L$-residue sequence $y=\{y_1, y_2, ..., y_L\} \in \mathcal{A}^L$ where $\mathcal{A}$ is the set of 20 common amino acids and $N \gg L$. We associate the biochemical features of each vertex to its nearest atom belonging to one of the $L$ residues. Specifically, we utilize two biochemical features for each vertex $\boldsymbol{x_i}$, which are its hydrophobicity $t_i$ and charge $c_i$. 
Then we sort all vertices based on the residue index of their nearest atoms.  Appendix Figure~\ref{Fig: surface} (a) depicts an example of raw surface.

\textbf{Surface Smoothing.}
As mentioned in previous methods~\cite{alexa2001point,lv2021voxel}, raw point clouds generally carry noise, which may limit the expressivity of the molecular surface. Therefore, point cloud denoising and smoothing are necessary. We apply Gaussian kernel smoothing on raw point cloud data:
\begin{equation}
\small 
\label{equation1}
\boldsymbol{x}_i^{'} =  \sum_{\boldsymbol{x}_j\in \mathcal{N}(\boldsymbol{x}_i)} \frac{\mathcal{K}(\boldsymbol{x}_i, \boldsymbol{x}_j)\boldsymbol{x}_j}{\sum_{\boldsymbol{x}_t\in \mathcal{N}(\boldsymbol{x}_i)}\mathcal{K}(\boldsymbol{x}_i, \boldsymbol{x}_t)}, \quad \mathcal{K}(x,y)=e^{-\frac{(x-y)^2}{\eta}}
\end{equation}
where $i\in \{1, 2, ..., N\}$. $\boldsymbol{x}_i$ and $\boldsymbol{x}_j$ denote the coordinates of $i$-th and $j$-th vertices on the raw point cloud, respectively. $\mathcal{N}(\boldsymbol{x}_i)$ are $K$-nearest neighbors of $\boldsymbol{x}_i$. 
$\mathcal{K}(\cdot,\cdot)$ is the Gaussian kernel with $\eta$ indicating distance scale in the point space.
In our paper, we set $\eta=\max(\mathrm{dist}(\boldsymbol{x}_i, \mathcal{N}(\boldsymbol{x}_i)))$ where $i\in \{1,...,N\}$. The number of nearest neighbors $K$ is set to 8. 
An in-depth analysis of kernel smoothing supports that the surface is infinitely smooth, i.e., $\theta \in C^{\infty}$~\cite{levin1998approximation,levin2004mesh}. Appendix Figure~\ref{Fig: surface} (b) depicts an example of smoothed surface.

\textbf{Surface Compression.} To reduce surface points and improve sample efficiency, we use an octree-based compression method to down-sample a protein surface~\cite{schnabel2006octree}. We use an octree to convert the surface into small cubes and estimate local densities of each cube. Every octree node is recursively divided into eight equal octants. After each division, the number of points in each node is examined to determine whether or not to continue dividing the current node. The cubes with fewer points than the specific threshold $N_{\text{min}}$ are taken as leaf nodes and not divided further. After all the nodes are processed, a point cloud is converted into a number of unequal-volume cubes based on the point distribution. Lower density regions result in larger cubes. The desired number of points for each cube is $N_s = V_s * r$, where $V_s$ is the number of points in the $s$-th cube and $r$ is the desired down-sampling ratio. Figure~\ref{Fig: surface} (c) depicts an example of compressed surface.


\subsection{Hierarchical Surface Encoder}
\label{hierarchical_encoder}
We design a hierarchical encoder to model the geometric shape and biochemical properties of a protein surface.

\textbf{Local Perspective Modeling.}
Residues nearing each other exhibit strong interactions. To model such interactions among nearest vertices on a surface, we design a variant of equivariant graph convolutional layer~(EGCL) proposed by~\citet{satorras2021n} to capture local geometric and biochemical features~(Figure~\ref{Fig: model} (b) left module). Specifically, after surface compression, the surface has $N^{'}$ vertices~($N^{'}\le N$), each of which has a 3D coordinate $\boldsymbol{x}_i^{'}\in \mathbb{R}^3$ and two biochemical features $\boldsymbol{h}_i=[t_i, c_i]^T$ where $t_i$ denotes its hydrophobicity, $c_i$ denotes its charge and $i \in \{1, ..., N^{'}\}$. We calculate local messages as:
\begin{equation}
\small 
\begin{split}
\boldsymbol{m}_{ij}^{'} &= \mathrm{SiLU}(\phi_e([(\boldsymbol{h}_i^{l};\boldsymbol{h}_j^{l};||\boldsymbol{x}_i^{'}-\boldsymbol{x}^{'}_{j}||_2)])) \\
w_{ij}^{l} &= \frac{\exp (W_s^l \boldsymbol{m}_{ij}^{'} + b_s^l)}{\sum_{k\in \mathcal{N}(\boldsymbol{x}_i)} \exp (W_s^l \boldsymbol{m}_{ik}^{'} + b_s^l)}\\
\boldsymbol{m}_{ij}^{l+1} &= w_{ij}^{l} * \boldsymbol{m}_{ij}^{'}\\
\end{split}
\end{equation}
where vertex $j\in \mathcal{N}(\boldsymbol{x}_i)$ belongs to the $K$-nearest neighbors of vertex $i$. We set $K=30$ here. $\boldsymbol{h}^1_i=W_m \boldsymbol{h}_i$ where $W_m\in \mathbb{R}^{256\times 2}$ is a mapping matrix and $l=$ 1 to $L_l$ is the layer number for the local perspective modeling module. $W_s^l \in \mathbb{R}^{1\times 256}$ and $b^l_s \in \mathbb{R}$ are learnable parameters. [;] denotes concatenation operation. $\phi_e$ denotes multi-layer perceptron~(MLP). SiLU denotes SiLU activation function~\cite{elfwing2018sigmoid}. For each vertex, we propagate messages from its neighbors to update the node feature:
\begin{equation}
\small 
\begin{split}
\boldsymbol{c}_{i}^{l+1} &= \sum\nolimits_{j\in \mathcal{N}(\boldsymbol{x}_i)} \boldsymbol{m}_{ij}^{l+1} \\
\boldsymbol{h}_{i}^{l+1} &= \boldsymbol{h}_{i}^{l} + \mathrm{gate}(\boldsymbol{c}_{i}^{l+1}) \odot \boldsymbol{c}_{i}^{l+1}\\
\end{split}
\end{equation}
where $\mathrm{gate}$ is a gating mechanism achieved by a MLP followed by a sigmoid function, which is used to control the information flow over the local geometric shapes.

\textbf{Global Landscape Modeling.} To facilitate message passing over the whole desired surface, we design a global landscape encoder called FAMHA~(Figure~\ref{Fig: model} (b) right module). Its key idea is to incorporate the frame averaging technique~(FA, \cite{puny2021frame}) into a multi-head attention layer. 
The resulting operation not only enables the global biochemical features to spread out but also guarantees its SE(3) equivariance. 
Specifically, from the compressed point cloud $\boldsymbol{X}^{'}\in \mathbb{R}^{N^{'}\times 3}$, we calculate three principle components $\boldsymbol{v}_1$, $\boldsymbol{v}_2$, $\boldsymbol{v}_3 \in \mathbb{R}^3$ through principle component analysis~(PCA). With these three base coordinates, we define a frame $\mathcal{F}(\boldsymbol{X}^{'})$ as a function:
\begin{equation}
\small 
\mathcal{F}(\boldsymbol{X}^{'}) = \{([\alpha_1\boldsymbol{v}_1, \alpha_2\boldsymbol{v}_2, \alpha_3\boldsymbol{v}_3], \boldsymbol{t}) | \alpha_i\in \{-1, +1\}\}
\end{equation}
where $\boldsymbol{t}$ is the centroid of $\boldsymbol{X}^{'}$. 
The frame function forms an algebraic group of eight transformations.
We calculate the global message passing as follows:
\begin{equation}
\small 
\label{equation_global}
\widetilde{\boldsymbol{H}}=\frac{1}{|\mathcal{F}(\boldsymbol{X}^{'})|}\sum_{g\in \mathcal{F}(\boldsymbol{X}^{'})} \text{FAMHA}(\boldsymbol{H}^{L_l+1};g^{-1}\boldsymbol{X}^{'})
\end{equation}
where $\boldsymbol{H}^{L_l+1}=[\boldsymbol{h}_1^{L_l+1},...,\boldsymbol{h}_{N^{'}}^{L_l+1}]^T \in \mathbb{R}^{N^{'}\times 256}$ is the output vertex features from the local perspective modeling. $g^{-1}\boldsymbol{X}^{'}$ denotes translating $\boldsymbol{X}^{'}$ with $\boldsymbol{t}$ and rotating ${\boldsymbol{X}^{'}}$ with rotation matrix $[\alpha_1\boldsymbol{v}_1, \alpha_2\boldsymbol{v}_2, \alpha_3\boldsymbol{v}_3]$ for $g\in \mathcal{F}(\boldsymbol{X}^{'})$.
FAMHA is composed of $L_g$ stacked multi-head attention~(MHA) sub-layers and fully connected feed-forward networks~(FFN). A residue connection and a layer normalization are performed after each of the two sub-layers. Accordingly, the FAMHA can be formulated as follows:
\begin{equation}
\small 
\begin{split}
\boldsymbol{h}_i^{l+1}&=\mathrm{LayerNorm} \left(\mathrm{FFN}(\Tilde{\boldsymbol{h}}_i^{l})+\Tilde{\boldsymbol{h}}_i^{l} \right), \\ \Tilde{\boldsymbol{h}}_i^{l}&=\mathrm{LayerNorm}\left(\mathrm{MHA}(\boldsymbol{h}_i^l, \boldsymbol{H}_g^l)+\boldsymbol{h}_i^l \right)
\end{split}
\end{equation}
where $l\in \{1,..., L_g\}$ and $i\in\{1,...,N^{'}\}$. $\boldsymbol{h}_i^1=[\boldsymbol{h}_i^{L_l+1};g^{-1}\boldsymbol{X^{'}}]$ for $g\in \mathcal{F}(\boldsymbol{X}^{'})$. $\boldsymbol{H}_g^l=[\boldsymbol{h}_1^l, ..., \boldsymbol{h}_{N'}^l]^T$.


\subsection{Autoregressive Protein Decoder}
\label{surface_decoder}

Given the hidden representations encoding both geometric shapes and biochemical features, we use an autoregressive Transformer decoder~(i.e., GPT, \cite{vaswani2017attention}) to generate the protein sequence for a given surface:
\begin{equation}
\small
p(y_t)=\mathrm{TransDec}(y_{<t}, \widetilde{\boldsymbol{H}};\theta_{dec})
\end{equation}
where $p(y_t)$ is the probability of $t$-th residue in the protein sequence and  $\theta_{dec}$ denotes the learnable parameters. 

We minimize the negative log likelihood to train the overall model:
\begin{equation}
\small
\mathcal{L} = \sum\limits_{t=1}^L - \log p(y_t;\theta)
\end{equation}
where $\theta=\{\theta_{enc}, \theta_{dec}\}$ denotes the parameter set of our hierarchical encoder and autoregressive protein decoder.

\section{Experiments}
\label{experiments}
In this section, we first describe our implementation details in Section~\ref{implementation_detals}. Then we conduct extensive experiments and evaluate our proposed \model on one general protein design task, \textbf{Inverse Folding}~(Section~\ref{inverse_folding}) and two functional protein design tasks, \textbf{Binder Design}~(Section~\ref{binder_design}) and \textbf{Enzyme Design}~(Section~\ref{enzyme_design}). The specific experimental settings are introduced in each task section.

\subsection{Implementation Details}
\label{implementation_detals}
We set a maximum limit of $5,000$ vertices for each surface. Surfaces with fewer than $5,000$ vertices remain unchanged, while those exceeding this limit are compressed with a down-sampling ratio $r$ set to $5,000/N$, where $N$ denotes the original vertex count. The minimum vertex number in a cube $N_{\text{min}}$ in surface compression is set to 32. Local perspective modeling utilizes three layers, and global landscape modeling employs a two-layer FAMHA. The two biochemical features are mapped to a hidden space with a dimensionality of $256$.  The autoregressive decoder is built with $3$-layer Transformer decoder. The mini-batch size and learning rate are set to $4,096$ tokens and $5$e-$4$, respectively. The model, trained with one NVIDIA RTX A$6000$ GPU card, utilizes the Adam optimizer~\cite{kingma2014adam}. The detailed values for biochemical features are provided in Appendix Table~\ref{Tab: append_chem_feature}.

\subsection{Inverse Folding}
\label{inverse_folding}
This task is to design protein sequences that fold into given backbone structures. In our method, we design a protein sequence from a coarser structure -- a protein surface, instead of a rigid backbone structure. 

\textbf{Datasets.} Following previous work~\cite{dauparas2022robust,gao2022pifold}, we use the CATH 4.2 dataset curated by \citet{ingraham2019generative} and follow the same data splits of \citet{jing2020learning}. Due to the occasional failures in raw surface construction by the MSMS tool, we filter out these instances as well as proteins longer than $1,024$ residues. As a consequence, the training, validation, and test splits consist of $14525$, $468$, and $887$ samples, respectively. We use the same data splits for all models rigorously for a fair comparison. The vertex count statistics for the curated CATH 4.2 dataset is provided in Appendix Table~\ref{Tab: vertex_count}.

\textbf{Baseline Models.} We compare with the following baseline models: 
(1) \textbf{ProteinMPNN} is a representative inverse folding model. (2) \textbf{PiFold} and  (3) \textbf{LM-DESIGN} are state-of-the-art-methods for inverse folding task. The used architecture for LM-DESIGN is LM-DESIGN (pretrained ProteinMPNN-CMLM: fine-tune). We use all their released codes on GitHub and the experimental settings in their official implementations to ensure a fair comparison. 

\textbf{Evaluation Metrics.} Following previous work~\cite{jing2020learning,gao2022pifold}, we use \textbf{perplexity} and \textbf{recovery rate} to evaluate the quality of designed protein sequences. Since the surface does not include residues buried beneath it, we report the recovery rate after pairwise alignment for all autoregressive models to ensure a fair comparison:
\begin{equation}
\small
\text{recovery rate}=\frac{\text{number of recovered residue}}{\text{aligned sequence length}}
\end{equation}
We provide the recovery rates after pairwise alignment for all models in Appendix Table~\ref{table_inverse_folding_appendix}, wherein the non-autoregressive models consistently exhibit lower recovery rates compared to the ones before alignment.

\begin{table}[!t]
\small
\begin{center}
\begin{tabular}{lcc}
\midrule
Methods  & Perplexity ($\downarrow$) & Recovery Rate~(\%, $\uparrow$)  \\
\midrule
ProteinMPNN & $5.19$ & $44.95$ \\
PiFold & $4.88$ & $52.61$ \\
LM-DESIGN & $4.47$ & $54.16$ \\
\rowcolor{myblue}
\model & $\textbf{3.13}$ & $\textbf{57.78}$ \\
\bottomrule
\end{tabular}
\end{center}
\caption{Perplexity and Recovery Rate of different approaches on CATH 4.2 dataset. ($\uparrow$): the
higher the better. ($\downarrow$): the lower the better. Among all the baselines, \model achieves the highest recovery rate and the lowest perplexity.}
\label{table_inverse_folding}
\end{table}

\begin{table*}[!t]
\small
\begin{center}
\begin{tabular}{lccccccc}
\midrule
\multirow{2}{*}{Models} & \multicolumn{3}{c}{Seen Class} & \multicolumn{3}{c}{Zero-Shot} & \multirow{2}{*}{Average} \\
\cmidrule(r){2-4} \cmidrule(r){5-7} 
& InsulinR  & PDGFR & TGFb & H3 & IL7Ra & TrkA &  \\
\midrule
Positive Binder & \textbf{5.9996} & \textbf{14.1366} & \textbf{15.4884} & \textbf{21.2631} & \textbf{20.9102} & \textbf{10.2791} & \textbf{14.7061} \\
Negative Binder & 19.7167 & 18.0937 & 23.2664 & 22.4556 & 26.0540 &24.7567 & 21.1335\\
\hdashline
Random Baseline & 19.9880 & 21.2690 & 21.4971 & 24.4997 & 24.1541 & 23.1147 &22.2020\\
ProteinMPNN~\cite{dauparas2022robust} & 18.3393 & 25.2919 & 25.8559 & 24.5968 & 25.5278 & 27.0980 & 23.4462 \\
PiFold~\cite{gao2022pifold} & 12.9809 & 21.8230 &24.4737 &23.3924 &26.6738  & 19.7172 & 20.5785\\
LM-DESIGN~\cite{zheng2023structure} & 13.6440 &  22.0749 &23.3725 & 23.8332 &24.3937 & 22.3987 &20.7728\\
\rowcolor{myblue}
\model & \textbf{10.2608} & \textbf{17.9862} & \textbf{17.7364} & \textbf{21.2916} & \textbf{20.8594} & \textbf{10.6535} &\textbf{16.9485}\\
\hdashline
\rowcolor{myblue}
\model-Pretrain & 11.2530& 18.4141 & \textbf{15.4011} & 22.2704 & \textbf{20.5700} & 21.3515 & 17.6699\\
\bottomrule
\end{tabular}
\end{center}
\caption{AF2 pAE\_interaction ($\downarrow$) for all models in the binder design task. ``Average" denotes the average AF2 pAE\_interaction across the entire test set instead of the direct average on different target proteins. We also provide the AF2 pAE\_interaction for randomly sampled negative binders of the same length as the positive ones. Our \model outperforms all previous methods on AF2 pAE\_interaction.}
\label{table_binder_greedy}
\end{table*}

\begin{table*}[!t]
\small
\begin{center}
\begin{tabular}{lccccccc}
\midrule
\multirow{2}{*}{Models} & \multicolumn{3}{c}{Seen Class} & \multicolumn{3}{c}{Zero-Shot} & \multirow{2}{*}{Average} \\
\cmidrule(r){2-4} \cmidrule(r){5-7} 
& InsulinR  & PDGFR & TGFb & H3 & IL7Ra & TrkA &  \\
\midrule
ProteinMPNN~\cite{dauparas2022robust} & 3.22 & 5.71 & \textbf{20.71} & 18.68& 24.10 & 7.50 & 11.96\\
PiFold~\cite{gao2022pifold} & 20.64 & 3.57 & 19.19 & \textbf{29.21} & 22.85 & 20.00 & 19.32\\
LM-DESIGN~\cite{zheng2023structure} & 7.74 & 15.00 & 15.71 & 22.29 & \textbf{24.28} & \textbf{25.00} & 16.37\\
\rowcolor{myblue}
\model & \textbf{31.57} & \textbf{19.99} & 11.61 & 23.21 & 19.28 & \textbf{25.00} & \textbf{22.29}\\
\hdashline
\rowcolor{myblue}
\model-Pretrain & 5.48 & \textbf{27.14} & \textbf{33.57} & \textbf{37.63} & \textbf{38.57} & \textbf{25.00} & \textbf{26.22} \\
\bottomrule
\end{tabular}
\end{center}
\caption{Success rate (\textbf{\%}, $\uparrow$)  for all models in the binder design task. ``Average" denotes the average success rate across the entire test set instead of the direct average on different target proteins. Our \model-Pretrain outperforms all previous methods in success rate by a big margin.}
\label{table_binder_sampling}
\end{table*}

\textbf{Main Results.} \textbf{Table~\ref{table_inverse_folding} shows that \model achieves the highest recovery rate and the lowest perplexity among all the compared baselines.} These findings demonstrate that incorporating both geometric and biochemical constraints of protein surfaces is beneficial for general protein design, leading \model to achieve the highest recovery rate across diverse protein folds in CATH 4.2 dataset.

\subsection{Protein Binder Design}
\label{binder_design}
In this section, we aim to use \model to design proteins that bind to a target protein with high affinity.

\textbf{Function Evaluator.} 
Following previous work~\cite{bennett2023improving,watson2023novo}, we use the \textbf{AlphaFold2}~(AF2) \textbf{pAE\_interaction} to evaluate the binding affinities between the designed binders and target proteins. 
\citet{bennett2023improving} discover that AF2 pAE\_interaction is very effective in distinguishing experimentally validated binders from non-binders, with a success rate ranging from 1.5\% to 7\% on target protein IL7Ra, TrkA, InsulinR and PDGFR.
We use the official code of \citet{bennett2023improving} to calculate the AF2 pAE\_interaction. The lower the AF2 pAE\_interaction, the better the designed binder. 

\textbf{Evaluation Metrics.} 
We calculate the average \textbf{AF2 pAE\_interaction} for the entire test set using greedy decoding, and the average \textbf{success rate}. For each positive binder, the success rate is defined as the proportion of designed binders with a lower pAE\_interaction than it.
For each $<$positive binder, target protein$>$ pair, we generate $10$ new binders using sampling with a temperature=$0.1$.
To calculate the pAE\_interaction, we first use ESMFold~\cite{lin2023evolutionary} to predict the structure of the designed binder sequence, and then we superimpose this structure to the real complex. 
Finally we calculate the AF2 pAE\_interaction for the new complex. 
As the AF2 pAE\_interaction model will automatically revise the input complex structures, there is little difference between AlphaFold2 and ESMFold predicted binder structures. A comparison is provided in Appendix Table~\ref{table_binder_af2}.

\textbf{Datasets.} We collect experimentally confirmed positive complexes of $<$binder, target protein$>$ pairs across six categories from \citet{bennett2023improving}. Among the 10 categories available, 4 exhibit indistinguishable AF2 pAE\_interaction between negative and positive binders. Therefore, we choose to evaluate on the remaining 6 categories for a reliable evaluation. For categories with over $50$ complexes, we employ an $8:1:1$ random split for training, validation, and test sets; otherwise, all complexes are included in the test set, establishing a zero-shot scenario. Detailed data statistics is provided in Appendix Table~\ref{table_binder_statistics}.

\textbf{Baseline Models.} Using the binder design dataset, we finetune all baseline models (ProteinMPNN, PiFold, LM-DESIGN) and our \model that are respectively pretrained on the CATH 4.2 dataset as detailed in Section~\ref{inverse_folding}. In addition, we also provide a random baseline by randomly mutating 20\% residues of a binder. 
Furthermore, to leverage all available crystal structures and fully explore the design capability of our \model, we pretrain the model using all surfaces generated from the entire Protein Data Bank~(PDB). This pretraining dataset, collected up to March 10, 2023, comprises $179,278$ $<$surface, sequence$>$ pairs.  Detailed data preprocessing steps and pretraining details are provided in the Appendix~\ref{pdb_data_collection}. We then finetune this model on the binder design dataset, and we refer to the resulting model as \model-Pretrain.

\textbf{Main Results.} The AF2 pAE\_interaction and success rate results of different models are reported in Table~\ref{table_binder_greedy} and Table~\ref{table_binder_sampling}, respectively. \textbf{The results show that our \model achieves the lowest average AF2 pAE\_interaction and the highest average success rate across six target proteins.} In particular, our \model achieves the best pAE\_interaction on all six categories and the highest success rate in three out of six categories. The pAE\_interaction of our model is even slightly lower than the experimentally confirmed positive binders on IL7Ra.
These findings demonstrate that leveraging protein surface properties is effective for functional binder design. 
Furthermore, \model achieves the highest success rates in two zero-shot testing categories, affirming its capability to directly capture valuable protein properties from surfaces. Consequently, even without specific training on binder proteins binding to particular targets, \model can generate binders in those categories with lower pAE\_interaction than the positive ones.
After pretraining on the entire PDB, the functions of binders generated by greedy decoding show minor difference. However, the success rate is improved significantly, from $22.29\%$ to $26.22\%$. It demonstrates that pretraining on larger dataset helps to improve the design capability of our \model, ensuring that more binders with better pAE\_interaction can be designed.

\begin{table*}[!t]
\small
\begin{center}
\begin{tabular}{lcccccc}
\midrule
\multirow{2}{*}{Models} & \multicolumn{4}{c}{Seen Class} & Zero-Shot & \multirow{2}{*}{Average} \\
\cmidrule(r){2-5} \cmidrule(r){6-6} 
& C00002  & C00677 & C00019 & C00003 & C00001 &  \\
\midrule
Real Enzyme &0.9573 & 0.8642 &0.4497 &0.8076 &0.9892 &0.9091\\
\hdashline
Random Baseline & 0.5523 &0.2475& 0.1673 &0.4705 &0.7891& 0.5292 \\
ProteinMPNN~\cite{dauparas2022robust} & \textbf{0.9711} &0.7375 &0.2614 &0.6699 &0.9763 &0.8676\\
PiFold~\cite{gao2022pifold} &0.9142 &0.8816 & 0.4296 &\textbf{0.8212} &0.9616 &0.8865\\
LM-DESIGN~\cite{zheng2023structure} & 0.9498 &0.8836 & \textbf{0.4585} & 0.8078 &0.9650 & \textbf{0.9037}\\
\rowcolor{myblue}
\model &0.9264 & \textbf{0.8921} &0.3892 & 0.7631 & \textbf{0.9772} &
0.8931\\
\hdashline
\rowcolor{myblue}
\model-Pretrain & \textbf{0.9376} & 0.8631 & \textbf{0.3949} & \textbf{0.7668} &0.9691 & 0.8900 \\
\bottomrule
\end{tabular}
\end{center}
\caption{ESP score ($\uparrow$) for all models by greedy decoding in enzyme design task. The substrate is denoted as its KEGG database ID. ``Average" here denotes the average ESP score across the entire test set instead of the direct average on different enzyme categories.}
\label{table_enzyme_greedy}
\end{table*}

\begin{table*}[!t]
\small
\begin{center}
\begin{tabular}{lcccccc}
\midrule
\multirow{2}{*}{Models} & \multicolumn{4}{c}{Seen Class} & Zero-Shot & \multirow{2}{*}{Average} \\
\cmidrule(r){2-5} \cmidrule(r){6-6} 
& C00002  & C00677 & C00019 & C00003 & C00001 &  \\
\midrule
ProteinMPNN~\cite{dauparas2022robust} & 47.54 & 31.63 &58.82 & 44.72 & 27.65 & 39.23\\
PiFold~\cite{gao2022pifold} & \textbf{48.54} & 41.72 & 58.29 & 37.54 & 24.97 & 40.65\\
LM-DESIGN~\cite{zheng2023structure} & 45.00 & 42.54 & 43.76 & \textbf{53.63} & 20.13 & 37.58\\
\rowcolor{myblue}
\model & 43.36 & \textbf{46.00} & \textbf{59.41} & 45.45 & \textbf{33.55} & \textbf{42.23}\\
\hdashline
\rowcolor{myblue}
\model-Pretrain & \textbf{50.90} & 41.81 & 52.94 & 36.36 & \textbf{34.21} & \textbf{43.63} \\
\bottomrule
\end{tabular}
\end{center}
\caption{Success rate (\textbf{\%}, $\uparrow$) for all models in the enzyme design task. ``Average" denotes the average success rate across the entire test set instead of the direct average on different enzyme categories. The substrate is denoted as its KEGG database ID. Notice that our \model achieves the highest average success rate.}
\label{table_enzyme_sampling}
\end{table*} 

\subsection{Enzyme Design}
\label{enzyme_design}
In our work, we target at designing enzymes which bind to specific substrates.

\textbf{Function Evaluator.} To evaluate the binding affinity between enzyme and substrate, we use the \textbf{ESP score} developed by \citet{kroll2023general}. Their model predicts enzyme-substrate interaction with 91\% accuracy across multiple benchmarks. We use their official code to calculate ESP score. 

\textbf{Evaluation Metrics.}  
Similar to binder design, we report the average \textbf{ESP score} using greedy decoding and average \textbf{success rate} using sampling with temperature=0.1.

\textbf{Datasets.} We collect five categories of enzymes from \citet{kroll2023general}, each of which binds to a specific substrate. We exclude enzymes in CATH 4.2 to prevent data leakage issue. For enzyme categories containing more than $100$ samples, we randomly split the data into training, validation, and test sets using an $8:1:1$ ratio after clustering; otherwise, all data are taken as the test set. The detailed data statistics is provided in Appendix Table~\ref{table_enzyme_statistics}.

\begin{table}[!t]
\small
\begin{center}
\begin{tabular}{lcc}
\midrule
Methods  & Perplexity ($\downarrow$) & Recovery Rate~(\%, $\uparrow$)  \\
\midrule
\rowcolor{myblue}
\model & $\textbf{3.13}$ & $\textbf{57.78}$ \\
~-- w five & 3.15 & 56.78 \\
~-- w/o global & 15.97 & 15.51 \\
~-- w/o local & 7.40 & 34.70 \\
~-- w hydrophobicity & 4.12 & 51.60 \\
~-- w charge & 11.23 & 21.34 \\
~-- w/o feature & 17.36 & 6.45 \\
~-- w unsorted & 18.98 & 14.95 \\
\bottomrule
\end{tabular}
\end{center}
\caption{Ablation study on CATH 4.2 dataset. ($\uparrow$): the higher the better. ($\downarrow$): the lower the better.}
\label{table_ablation_study}
\end{table}

\textbf{Baseline Models.} Similar to binder design, we finetune all baseline models (ProteinMPNN, PiFold, LM-DESIGN) and our \model, using the enzyme design dataset, on the pretrained models from inverse folding task, respectively. Likewise, we also provide the results of a random baseline and \model-Pretrain with the same settings as binder design. Note that the pretraining dataset excludes all enzymes here to prevent data leakage issue.

\textbf{Main Results.} Table~\ref{table_enzyme_greedy} and Table~\ref{table_enzyme_sampling} show that \textbf{our \model achieves the highest average success rate and comparable average ESP score to LM-DESIGN across five categories.} It is important to note that LM-DESIGN is finetuned from the 650M ESM-1b~\cite{rives2021biological}, which is pretrained on the extensive UniRef50 dataset. Consequently, there is a potential for data leakage, enabling it to achieve the best performance on average ESP score. 
However, our \model outperforms LM-Design with a significantly higher success rate of 42.23\% compared to LM-Design's 37.58\%. This performance is further improved to 43.63\% after pretraining on the entire PDB surfaces. These findings indicate that our \model is able to design enzymes with stronger enzyme-substrate interaction functions than real enzymes, validating that surface properties are helpful for functional protein design again. Furthermore, our \model demonstrates zero-shot design capability, displaying a success rate of 33.55\% for designing enzymes binding to substrate C00001.

\section{Analysis: Diving Deep into \model}
\label{analysis}
\subsection{Ablation Study: How Does Each Component Work?}
\textbf{Both geometric and biochemical constraints facilitate protein design.}
To better analyze the influence of different components in our model, we conduct ablation tests on inverse folding task. The models to be compared are listed as follows:
(1) \model-w-five uses five biochemical features, which are hydrophobicity, charge, polarity, acceptor and donor;
(2) \model-w/o-global removes the global landscape modeling;
(3) \model-w/o-local removes the local perspective modeling;
(4) \model-w-hydrophobicity only uses hydrophobicity feature;
(5) \model-w-charge only uses charge feature;
(6) \model-w/o-feature does not use any biochemical features;
(7) \model-w-unsorted does not sort the vertices on the raw surface.

The results in Table~\ref{table_ablation_study} indicate that incorporating five chemical features does not yield additional benefits. Removal of either global landscape modeling or local perspective modeling results in significant performance degradation. Utilizing only the hydrophobicity feature slightly decreases the performance, while relying solely on the charge feature seriously damages performance. The absence of both biochemical features further decreases performance. 
These observations validate the crucial roles played by both geometric shapes and biochemical features in surface representation learning, emphasizing the necessity of incorporating both into the protein design process. 
It is worth noting that unsorting the vertices on the raw surface significantly decreases performance. Our interpretation is the local shape in different areas might be similar, and the model is hard to align each local shape to a specific protein fragment without sorting the vertices, especially for extremely long sequences.

\begin{figure}
\begin{minipage}[t]{0.5\linewidth}
\centering
\includegraphics[width=3.9cm]{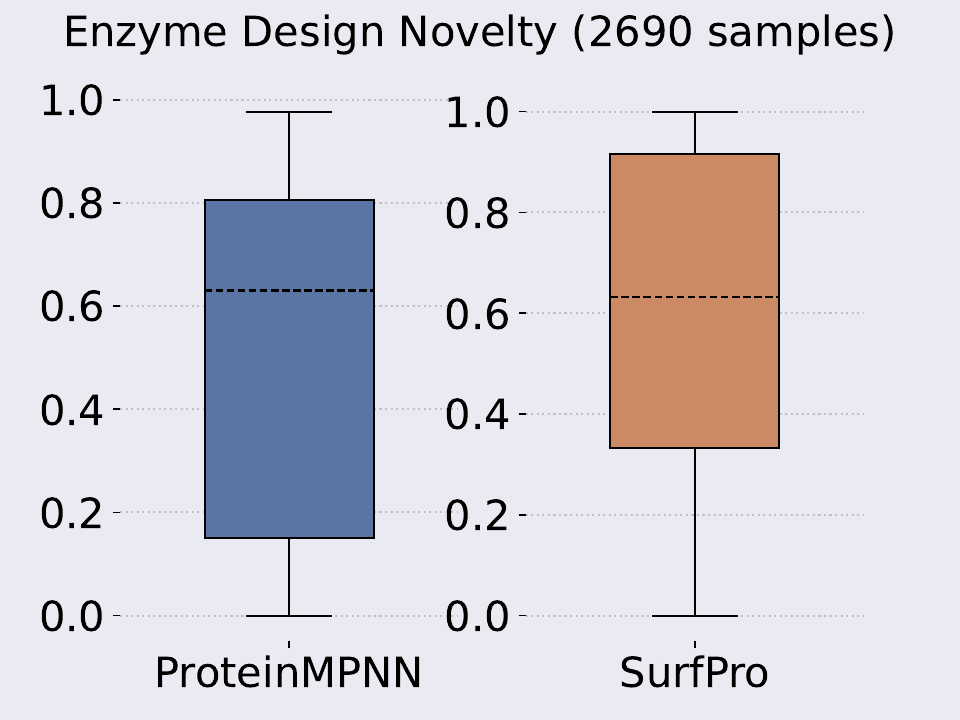}
\centerline{\small{(a) enzyme novelty}}
\end{minipage}%
\begin{minipage}[t]{0.5\linewidth}
\centering
\includegraphics[width=3.9cm]{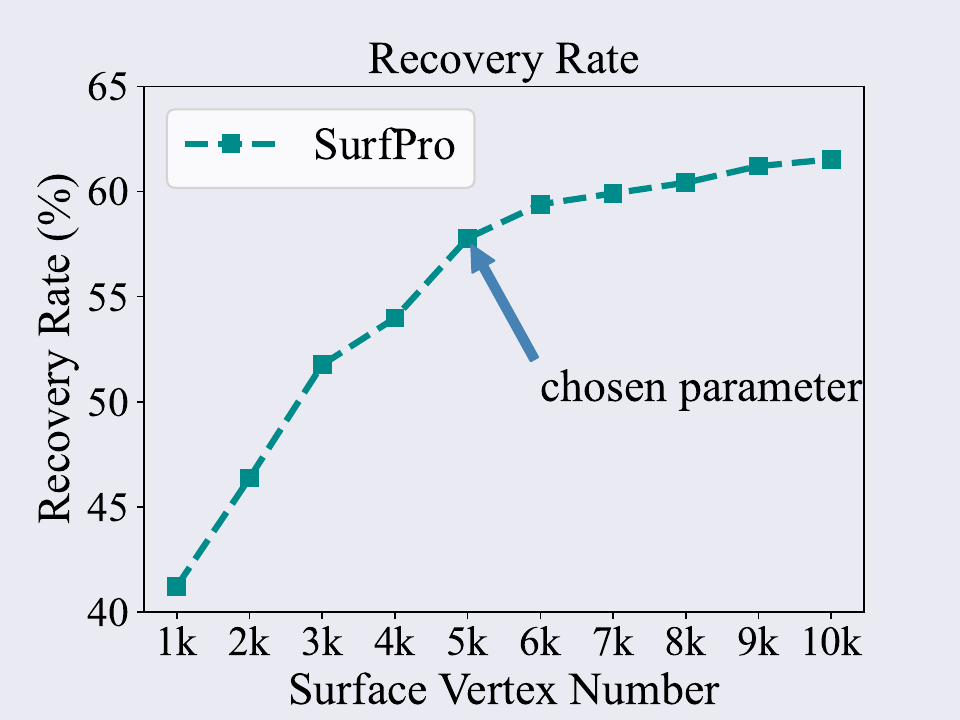}
\centerline{(b) surface vertex number effect}
\end{minipage}
	\caption{(a) Novelty~(1-recovery rate) of designed enzymes with temperature=0.1. (b) Recovery rates on CATH 4.2 dataset for models with varying numbers of sampled surface vertices.} 
 \label{Fig: Analysis_diversity_mutation}
\end{figure} 

\begin{table}[!t]
\small
\begin{center}
\begin{tabular}{lcc}
\midrule
Methods  & Perplexity ($\downarrow$) & Recovery Rate~(\%, $\uparrow$)  \\
\midrule
\rowcolor{myblue}
\model & $\textbf{3.13}$ & $\textbf{57.78}$ \\
~-- w NAD & 12.51 & 9.61 \\
\bottomrule
\end{tabular}
\end{center}
\caption{Ablation study on CATH 4.2 dataset. ($\uparrow$): the higher the better. ($\downarrow$): the lower the better. NAD denotes non-autoregressive decoder.}
\label{table_ablation_study_nat}
\end{table}

\subsection{Ablation Study: Can Autoregressive Decoder Be Replaced?}
\textbf{The autoregressive decoder demonstrates strong performance within our proposed \model architecture.} To assess its significance, we compare \model with an alternative employing a non-autoregressive decoder, incorporating SoftCopy and glancing learning strategies inspired by non-autoregressive machine translation in natural language processing~\cite{gu2021fully,qian2021glancing}. Table~\ref{table_ablation_study_nat} presents the comparative results. It unequivocally demonstrates that \model outperforms the non-autoregressive decoder variant, affirming the efficacy of the autoregressive decoder within our proposed framework.

\subsection{Can \model Design Novel and Diverse Proteins?}
\begin{figure}
  \centering
  \includegraphics[width=8.0cm]{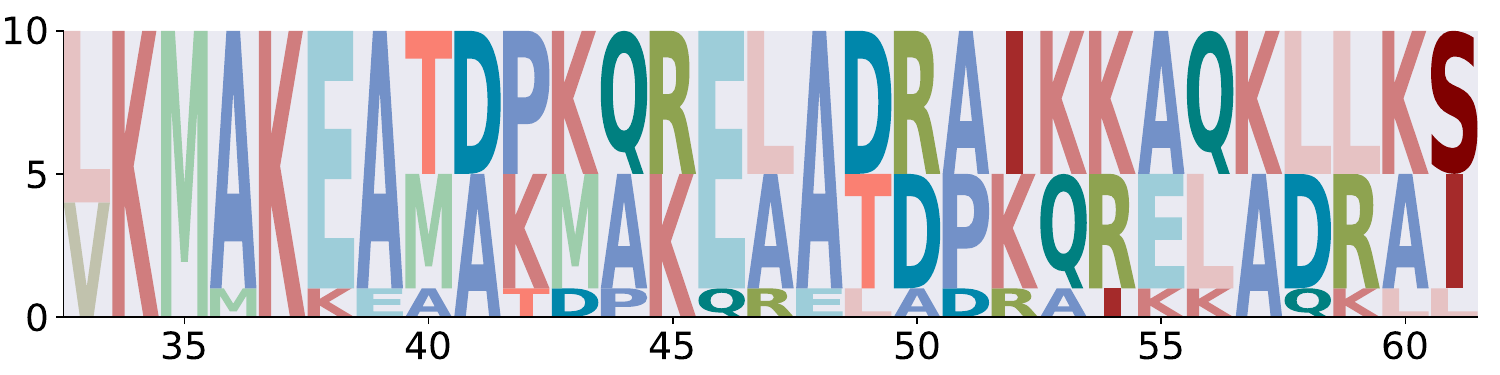}
  \caption{Amino acid distribution at different positions in designed binders for the target protein InsulinR.}
  \label{Fig: mutation}
\end{figure}

\textbf{\model is able to generate novel and diverse proteins.} The novelty distribution among the designed enzymes, sampled with a temperature of 0.1, is visualized in Figure~\ref{Fig: Analysis_diversity_mutation} (a). Novelty here is calculated as 1 - recovery rate. The figure shows our \model demonstrates a superior average novelty of $58.51\%$ in comparison to ProteinMPNN, which achieves an average novelty of $49.46\%$.
Additionally, we analyze the amino acid distribution in designed binders across different positions. An illustrative example is shown for target protein InsulinR in Figure~\ref{Fig: mutation}, demonstrating a diversified residue distribution. Notably, all these designed binders~(provided in Appendix~\ref{Appendix: designed_binders}) achieve a pAE\_interaction lower than 10, a threshold identified by \citet{bennett2023improving} to significantly enhance success rate. These findings affirm that \model is able to design diverse proteins with desired functions.

\subsection{How Does Vertex Number Affect Protein Design?}
\textbf{The more vertices sampled, the more effectively \model performs.}
To thoroughly explore the impact of surface vertex sampling size on model performance, we train models with maximum vertex numbers ranging from 1k to 10k on CATH 4.2 dataset. 
Figure~\ref{Fig: Analysis_diversity_mutation} (b) demonstrates an improved recovery rate as the number of sampled vertices increases. However, after reaching a sampling size exceeding 5k, the improvement rate slightly decreases. Additionally, with more vertices sampled, the inference speed will decrease. To ensure both design quality and inference efficiency, we set the maximum vertex number to 5k.

\begin{figure}
\begin{minipage}[t]{0.5\linewidth}
\centering
\includegraphics[width=4.0cm]{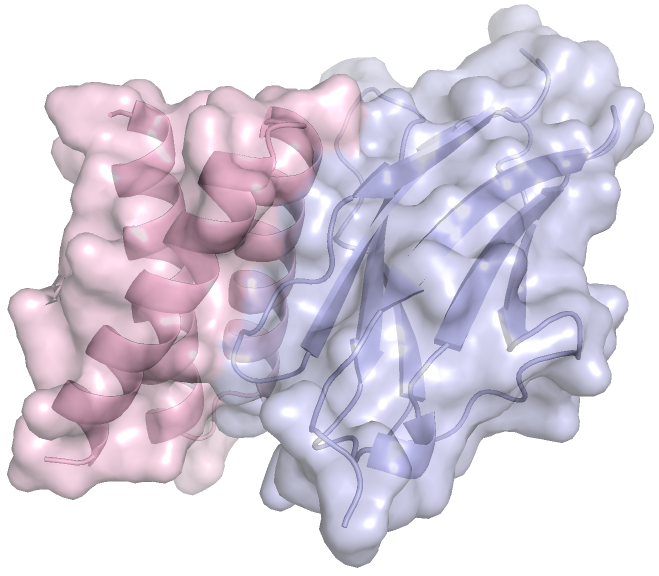}
\centerline{\small{(a) pAE score with TrkA: 4.75}}
\end{minipage}%
\begin{minipage}[t]{0.5\linewidth}
\centering
\includegraphics[width=4.3cm]{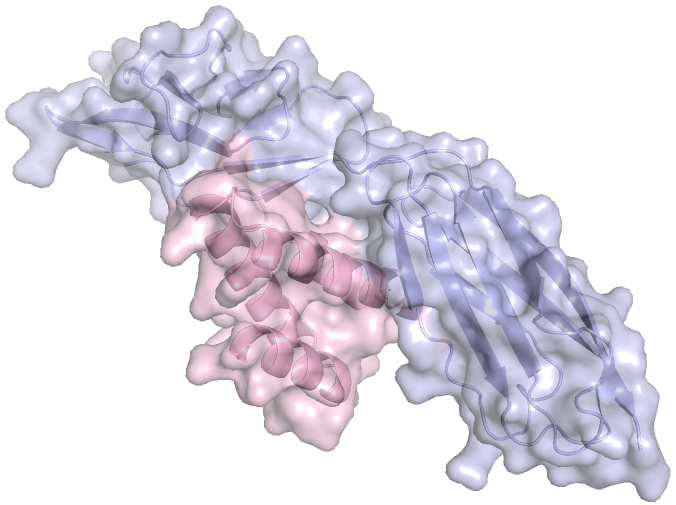}
\centerline{\small{(b) pAE score with PDGFR: 5.58}}
\end{minipage}
	\caption{Case study of complexes involving our \model designed binders (in red) and target proteins (in purple): (a) target protein TrkA with pAE\_interaction=4.75, (b) target protein PDGFR with pAE\_interaction=5.58.} 
 \label{Fig: Case_study}
\end{figure}

\begin{table*}[!t]
\small
\begin{center}
\begin{tabular}{lcccccccc}
\midrule
\multirow{2}{*}{Models} & \multicolumn{3}{c}{Seen Class} & \multicolumn{3}{c}{Zero-Shot} & \multirow{2}{*}{Average} & \multirow{2}{*}{Inference Time (s/sample)} \\
\cmidrule(r){2-4} \cmidrule(r){5-7} 
& InsulinR  & PDGFR & TGFb & H3 & IL7Ra & TrkA &  \\
\midrule
MaSIF & \textbf{9.0737} &  \textbf{10.8734}&18.8502&21.9093&25.9689&25.9112&\textbf{16.1280} & 210.42\\
\rowcolor{myblue}
\model & 10.2608 & 17.9862 & \textbf{17.7364} & \textbf{21.2916} & \textbf{20.8594} & \textbf{10.6535} &16.9485 & \textbf{0.45}\\
\bottomrule
\end{tabular}
\end{center}
\caption{Comparison with MaSIF on binder design task. ``Average" denotes the average AF2 pAE\_interaction across the entire test set instead of the direct average on different target proteins.}
\label{table_binder_masif}
\end{table*}

\subsection{Comparing with MaSIF}
In this section, we conduct a comparative analysis between our \model and the well-established surface-based binder design model MaSIF~\cite{gainza2020deciphering}. Initially, we employ ProteinMPNN to generate 100 binder sequences for each positive binder. Subsequently, ESMFold is utilized to predict the structures of these designed binder sequences. Following this, MaSIF is employed to rank these candidate binders, with the top-performing one selected as the final designed binder. The resulting pAE\_interaction scores across six datasets are summarized in Table~\ref{table_binder_masif}. The comparison reveals that MaSIF outperforms our model in 2 out of 6 target proteins, while our \model outperforms in 4 out of 6 target proteins. Despite MaSIF demonstrating a slightly superior average pAE\_interaction score, it's noteworthy that the average time taken for MaSIF to identify a promising binder is 210.42 seconds, significantly longer than the mere 0.45 seconds required by \model. This underscores \model's efficacy as a standout generative model, capable of swiftly and directly generating functional protein sequences in an end-to-end manner.

\subsection{Case Study}

To get an insight on the designed functional proteins by our \model, we visualize two complexes of our model designed binders and target proteins belonging to TrkA~(Figure~\ref{Fig: Case_study} (a)) and PDGFR~(Figure~\ref{Fig: Case_study} (b)). Both the complexes have AF2 pAE\_interaction lower than $6$, indicating a strong protein-protein binding. As \citet{bennett2023improving} state in their work that success rates for designs will be sharply increased with AF2 pAE\_interaction $<$ 10. It intuitively shows that our \model is able to design functional binders with high protein-protein binding affinities.

\section{Discussion}
Our \model exhibits superior performance in rapidly and directly generating functional protein sequences based on provided protein surfaces. However, despite its impressive capabilities, certain limitations persist, which we will discuss in this section. While our approach excels in protein optimization, it leans more towards refinement rather than de novo protein design. This distinction is significant and meaningful. Particularly in binder design, achieving a high-affinity binder from scratch is seldom feasible in practical scenarios. Thus, starting from existing positive binders can accelerate the design process. Nonetheless, this approach also imposes constraints on the practical utility of our method. Firstly, locating a positive initial point isn't always feasible. Secondly, commencing from a favorable starting point may result in limited improvements, as our \model necessitates consideration of both geometric and biochemical constraints. Our future work could explore methods for de novo designing protein surfaces. For example, integrating diffusion models to generate point clouds could substantially enhance the versatility and applicability of our existing framework.

\section{Conclusion}
\label{conclusion}
In this work, we propose \model, a new generative model to design functional proteins based on desired surface. \model incorporates a hierarchical encoder that progressively captures geometric and biochemical features, transitioning from a local perspective to a global landscape. Additionally, an autoregressive decoder is employed to generate a protein sequence based on the learned geometric and biochemical representations of the surface. Our approach consistently outperforms prior strong inverse folding methods on a general protein design benchmark CATH 4.2, with a sequence recovery rate of 57.78\% , and two functional protein design tasks,  with higher success rates in terms of protein-protein binding and enzyme-substrate interaction scores.

\section*{Acknowledgement}
This research is partially supported by the UC Santa Barbara Faculty Research Grant (to L.L.).
The authors thank the anonymous reviewers and Wenxian Shi, Siqi Ouyang, Danqing Wang, and Yujia Gao for their excellent suggestions.

\section*{Impact Statement}
This paper contributes to the design of functional proteins from their surfaces, thereby advancing the field of generative molecule design. Our work may have various societal implications, but we do not think they need to be specifically highlighted here.





\bibliography{icml2024}
\bibliographystyle{icml2024}

\newpage
\appendix
\onecolumn
\section*{Appendix}

\section{Surface Biochemical Features}
\label{surfce_chemical_feature}
The specific values for biochemical features are sourced from the ImMunoGeneTics information system, which are provided in Table~\ref{Tab: append_chem_feature}. In our paper, we utilize two biochemical features, which are hydrophobicity and charge. Additionally, we present the results of our model applying five biochemical features in Table~\ref{table_ablation_study}. The extra three biochemical features include polarity, acceptor and donor, with their respective values provided in Table~\ref{Tab: append_chem_feature}.

\begin{table*}[ht]
\small 
\centering
\begin{tabular}{lcc}
\midrule
Feature & Description & Value \\
\midrule
hydrophobicity & the hydrophobicity level of a residue,  & I: 4.5, V: 4.2, L: 3.8, F: 2.8, C: 2.5, M: 1.9, A: 1.8  \\
& the higher the hydrophobicity, & W: -0.9, G: -0.4, T: -0.7, S: -0.8, Y: -1.3, P: -1.6, H: -3.2\\
& the more hydrophobic the residue & N: -3.5, D: -3.5, Q: -3.5,  E: -3.5, K: -3.9, R: -4.5\\
\hdashline
charge & the charge value of a residue & R: 1, K: 1, D: -1, E: -1, H: 0.1, others: 0 \\
\midrule
polarity & polarity is a separation of  & R: 1, N: 1, D: 1, Q: 1, E: 1, H: 1, K: 1, S: 1, T: 1, Y: 1 \\
& electric charge leading to a molecule & others: 0 \\
\hdashline
acceptor & accepting electrons from another compound & D: 1, E: 1, N: 1, Q: 1, H: 1, S: 1, T: 1, Y: 1, others: 0 \\
\hdashline
donor & transferring electrons to another compound & R: 1, K: 1, W: 1, N: 1, Q: 1, H: 1, S: 1, T: 1, Y: 1, others: 0 \\
\bottomrule
\end{tabular}
\vspace{-0.5em}
\caption{Detailed values for biochemical features.}
\label{Tab: append_chem_feature}
\end{table*}

\section{Protein Surface Generation Example}
\label{Appendix: surface_example}
\begin{figure}[ht]
\begin{minipage}[t]{0.33\linewidth}
\centering
\includegraphics[width=2.5cm]{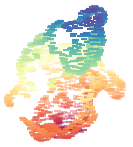}
\centerline{(a) raw surface}
\end{minipage}%
\begin{minipage}[t]{0.33\linewidth}
\centering
\includegraphics[width=2.4cm]{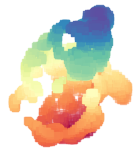}
\centerline{(b) smoothed surface}
\end{minipage}%
\begin{minipage}[t]{0.33\linewidth}
\centering
\includegraphics[width=2.52cm]{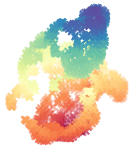}
\centerline{(c) compressed surface}
\end{minipage}
\vspace{-0.6em}
	\caption{Protein surface generation: (a) raw surface, (b) smoothed surface, and (c) compressed surface.} 
 \label{Fig: surface}
\end{figure}

We provide an example of protein surface generation process in Figure~\ref{Fig: surface}, including raw surface construction (Figure~\ref{Fig: surface} (a)), surface smoothing (Figure~\ref{Fig: surface} (b)) and surface compression (Figure~\ref{Fig: surface} (c)).

\section{Additional Information on Inverse Folding Task}
\subsection{Surface Vertex Count for CATH 4.2 Benchmark}
\label{surface_vertex_count}
Table~\ref{Tab: vertex_count} presents the vertex count statistics for the CATH 4.2 dataset.

\begin{table*}[ht]
\small 
\centering
\begin{tabular}{lccc}
\midrule
Vertex Count & Training & Validation & Test \\
\midrule
Average Vertex Count/Residue & $103$ & $114$ & $119$  \\
Maximum Vertex Count  & $529,271$ & $393,522$ & $385,008$ \\
\bottomrule
\end{tabular}
\vspace{-0.5em}
\caption{Vertex count statistics for surfaces from the CATH 4.2 dataset.}
\label{Tab: vertex_count}
\end{table*}

\subsection{Recovery Rate After Pairwise Alignment}

\begin{table}[!t]
\small
\begin{center}
\begin{tabular}{lc}
\midrule
Methods& Recovery Rate~(\%, $\uparrow$)  \\
\midrule
ProteinMPNN~\cite{dauparas2022robust} & $35.89$ \\
PiFold~\cite{gao2022pifold} & $39.29$ \\
LM-DESIGN~\cite{zheng2023structure} & $38.21$ \\
\rowcolor{myblue}
\model & $\textbf{57.78}$ \\
\bottomrule
\end{tabular}
\end{center}
\caption{Recovery rate after pairwise alignment of different approaches on CATH 4.2 dataset. ($\uparrow$): the
higher the better. Among all the baselines, \model achieves the highest recovery rate.}
\label{table_inverse_folding_appendix}
\end{table}

The amino acid recovery rates of different approaches after pairwise alignment are provided in Table~\ref{table_inverse_folding_appendix}. It shows our \model achieves the highest recovery rate among all the compared baselines. In addition, for all non-autoregressive models, the recovery rate decreases after pairwise alignment.

\section{Data Statistics for Two Functional Protein Design Tasks}
\label{Appendix: data_statistics}
\subsection{Binder Design Datasets}
The detailed data statistics for protein binder design is provided in Table~\ref{table_binder_statistics}.

\begin{table}[ht]
\small
\begin{center}
\begin{tabular}{lcccc}
\midrule
Target Protein  & Positive Binder & Training & Validation & Test  \\
\midrule
InsulinR & 238 & 184 & 23 & 31\\
PDGFR & 262 & 208 & 26 & 28 \\
TGFb & 95 & 72 & 9 & 14 \\
H3 & 38 & 0 & 0 & 38 \\
IL7Ra & 7 & 0 & 0 &7 \\
TrkA & 4 & 0 & 0 & 4 \\
Total & 644 & 464 & 58 & 122 \\ 
\bottomrule
\end{tabular}
\end{center}
\caption{Data statistics for the six categories of $<$positive binder, target protein$>$ complexes in the binder design task.}
\label{table_binder_statistics}
\end{table}

\subsection{Enzyme Design Datasets}
The detailed data statistics for enzyme design is provided in Table~\ref{table_enzyme_statistics}.

\begin{table}[ht]
\small
\begin{center}
\begin{tabular}{lcccc}
\midrule
Substrate  & Enzyme & Training & Validation & Test  \\
\midrule
C00002 & 1101 &	881	& 110 & 110\\
C00677 & 555	& 445 &	55	& 55 \\
C00019 & 175& 141 & 17 & 17 \\
C00003 & 112 & 90 & 11 & 11 \\
C00001 & 76	& 0 & 0	& 76 \\
Total & 1979 &	1557 & 193 & 269 \\ 
\bottomrule
\end{tabular}
\end{center}
\caption{Data statistics for the five enzyme categories, each of which binds to a specific substrate identified as its KEGG ID.}
\label{table_enzyme_statistics}
\end{table}

\section{PDB Surface Pretraining}
\label{pdb_data_collection}

To fully explore the design capability of our model, we pretrain our proposed \model on surfaces from the entire PDB. Specifically, we gather all proteins in the PDB until March 10, 2023, resulting in a total of 198,726 samples. Subsequently, we employ MSMS to compute the continuous surface for each protein and extract chain A as the target sequence. Following a similar data processing approach to the CATH 4.2 dataset, we filter out failed instances during the raw surface construction process by MSMS tool and sequences longer than 1024 residues. To prevent the potential data leakage issue, we exclude proteins included in the enzyme design datasets. Consequently, we obtain 179,278 $<$surface, sequence$>$ pairs. Among them, We randomly split 50 cases for the validation set, leaving the rest for the training set. The model undergoes pretraining for 1,000,000 steps. The learning rate and batch size are set to 5e-4 and 4096 tokens, respectively.
After pretraining process, we obtain \model-Pretrain.

Then we separately finetune \model-Pretrain on the binder design and enzyme design tasks, with the corresponding performance reported in Table~\ref{table_binder_greedy},~\ref{table_binder_sampling},~\ref{table_enzyme_greedy} and~\ref{table_enzyme_sampling}. The outcomes illustrate a notable improvement in the design capabilities of our \model, as evidenced by substantially increased success rates in both tasks. This enhancement implies that our model can effectively design proteins with stronger binding functions than experimentally confirmed positive binders or natural enzymes when pretrained on a larger dataset.

\section{Additional Experimental Results}
\subsection{Binder Design Success Rate Comparing AlphaFold2 and ESMFold Predicted Binder Structures}
\label{Appendix_binder_design_af2_emsfold}
We present the success rate for the binder design task, comparing the binder structures predicted by AlphaFold2 and ESMFold in Table~\ref{table_binder_af2}. The results reveal a minor difference of only $0.41\%$, indicating that there is little distinction in performance between AlphaFold2 and ESMFold for this task.

\begin{table*}[ht]
\small
\begin{center}
\begin{tabular}{lccccccc}
\midrule
\multirow{2}{*}{Models} & \multicolumn{3}{c}{Seen Class} & \multicolumn{3}{c}{Zero-Shot} & \multirow{2}{*}{Average} \\
\cmidrule(r){2-4} \cmidrule(r){5-7} 
& InsulinR  & PDGFR & TGFb & H3 & IL7Ra & TrkA &  \\
\midrule
\model-AlphaFold2 & \textbf{33.15} & 17.14 & 11.29 & 21.42 & 17.14 & \textbf{25.00} & 21.88 \\
\model-ESMFold & 31.57 & \textbf{19.99} & \textbf{11.61} & \textbf{23.21} & \textbf{19.28} & \textbf{25.00} & \textbf{22.29}\\
\bottomrule
\end{tabular}
\end{center}
\caption{Success rate (\textbf{\%})  for \model with AlphaFold2 predicted binder structures~(\model-AlphaFold2) and \model with ESMFold predicted binder structures~(\model-ESMFold) on the binder design task. ``Average" denotes the average success rate across the entire test set instead of the direct average on different target proteins.}
\label{table_binder_af2}
\end{table*}

\subsection{Sequences for Case Analysis}
\label{Appendix: designed_binders}
We provide the designed binder sequences and the positive one in Table~\ref{Tab: designed_binder_seqs} for cases analyzed in Figure~\ref{Fig: mutation} .

\begin{table*}[ht]
\small 
\centering
\begin{tabular}{lc}
\midrule
Binders & Sequence  \\
\midrule
positive one & DEFTEIVKELVKLAEEAVKKNDEDSVKFIEAMLKMMKEAATDPKQRELADRAIKKVQKLLKS\\
binder1 & DEFTEIVKELVKLAEEAVKKNDEESVKFIEAMVKMAKEAMAKMAKEAATDPKQRELADRAIK \\
binder2 & DEFTEIVKELVKLAEEAVKKNDEESVKFIEAMLKMAKEATDPKQRELADRAIKKAQKLLKS \\
binder3 & DEFTEIVKELVKLAEEAVKKNDEESVKFIEAMVKMAKEAMAKMAKEAATDPKQRELADRAIK\\
binder4 &DEFTEIVKELVKLAEEAVKKNDEESVKFIEAMLKMAKEATDPKQRELADRAIKKAQKLLKS \\
binder5 & DEFTEIVKELVKLAEEAVKKNDEESVKFIEAMVKMAKEAMAKMAKEAATDPKQRELADRAIK\\
binder6 & DEFTEIVKELVKLAEEAVKKNDEESVKFIEAMLKMAKEATDPKQRELADRAIKKAQKLLKS\\
binder7 & DEFTEIVKELVKLAEEAVKKNDEESVKFIEAMVKMAKEAMAKMAKEAATDPKQRELADRAIK\\
binder8 & DEFTEIVKELVKLAEEAVKKNDEESVKFIEAMLKMAKEATDPKQRELADRAIKKAQKLLKS\\
binder9 & DEFTEIVKELVKLAEEAVKKNDEESVKFIEAMLKMMKKEAATDPKQRELADRAIKKAQKLLK\\
binder10 & DEFTEIVKELVKLAEEAVKKNDEESVKFIEAMLKMAKEATDPKQRELADRAIKKAQKLLKS\\
\bottomrule
\end{tabular}
\vspace{-0.5em}
\caption{Sequences for cases analyzed in Figure~\ref{Fig: mutation}: designed binders and the experimentally confirmed positive one for the target protein InsulinR.}
\label{Tab: designed_binder_seqs}
\end{table*}

\end{document}